\documentclass[12pt]{article}
\usepackage{amsmath,amssymb}
\usepackage{graphics}
\usepackage[dvipdfm]{graphicx}

\begin{document}
\baselineskip=16pt
\begin{titlepage}
\begin{flushright}
\end{flushright}
\vspace*{1.2cm}

\begin{center}

{\Large\bf 
Why is TeV-scale a geometric mean of neutrino mass and GUT-scale?
} 
\lineskip .75em
\vskip 1.5cm

\normalsize
{\large Naoyuki Haba}

\vspace{1cm}

{\it Department of Physics, 
 Osaka University, Toyonaka, Osaka 560-0043, 
 Japan} \\

\vspace*{10mm}

{\bf Abstract}\\[5mm]
{\parbox{13cm}{\hspace{5mm}
%

Among three typical energy scales, 
 a neutrino mass scale ($m_\nu\sim$ 0.1 eV), 
 a GUT scale ($M_{GUT}\sim 10^{16}$ GeV), 
 and 
 a TeV-scale ($M_{NP}\sim 1$ TeV), 
 there is a fascinating relation of     
 $M_{NP}\simeq \sqrt{m_\nu\cdot M_{GUT}}$.  
The TeV-scale, $M_{NP}$, is a new physics
 scale beyond the standard model 
 which is regarded as 
 supersymmetry in this letter. 
We suggest 
 a simple supersymmetric neutrinophilic 
 Higgs doublet model, which 
 realizes the above relation dynamically as well as  
 the suitable $m_\nu$ through  
 a tiny vacuum expectation value 
 of neutrinophilic Higgs  
 without 
 additional scales other than 
 $M_{NP}$ and $M_{GUT}$.     
A gauge coupling unification, which is an excellent feature
 in the supersymmetric standard model, 
 is preserved 
 automatically 
 in this setup.

}}

\end{center}

\vspace{.5cm}
\hspace*{1cm}

\end{titlepage}

\section{Introduction}

There are 
 three typical energy scales, 
 a neutrino mass scale ($m_\nu\sim$ 0.1 eV), 
 a GUT scale ($M_{GUT}\sim 10^{16}$ GeV), 
 and  
 a TeV-scale ($M_{NP}\sim 1$ TeV)  
 which is a new physics 
 scale beyond the standard model (SM) and 
 regarded as 
 supersymmetry (SUSY) in this letter. 
Among these three scales, 
 we notice a fascinating relation,    
\begin{equation}
 M_{NP}^2 \simeq {m_\nu\cdot M_{GUT}}\ .  
\label{1}
\end{equation}
Is this relation an accident, or 
 providing a clue to the underlying 
 new physics ?
We take a positive stance toward 
 the latter possibility. 

As for a neutrino mass $m_\nu$, 
 its smallness is still a mystery, and 
 it is one of the
 most important clues to find new physics. 
Among a lot of possibilities, 
 a neutrinophilic Higgs doublet model 
 suggests an interesting explanation of 
 the smallness 
 by 
 a tiny vacuum expectation value (VEV) \cite{Ma}-\cite{Haba:2011fn}. 
This VEV from 
 a neutrinophilic Higgs doublet 
 is of ${\mathcal O}(0.1)$ eV 
 which is the same as the neutrino mass,
 so that it 
 suggests Dirac 
 neutrino\cite{Nandi,WWY,Davidson:2009ha,Davidson:2010sf}.\footnote{
In Refs.\cite{Ma,MaRa,Ma:2006km,HabaHirotsu,HabaTsumura,HS1,HS2}, 
 Majorana neutrino scenario is considered 
 through TeV-scale seesaw with  
 a neutrinophilic Higgs VEV of ${\mathcal O}(1)$ MeV.}  
Thus, the neutrino mass is much smaller than other fermions, 
 since its origin is 
 the tiny VEV from the different
 (neutrinophilic) Higgs doublet. 
Introduction of $Z_2$-symmetry 
 distinguishes the neutrinophilic Higgs from 
 the SM-like Higgs, 
 where $m_\nu$ is surely generated only through 
 the VEV of the neutrinophilic Higgs. 
The SUSY extension of the neutrinophilic doublet model 
 is considered in Refs.\cite{{Marshall:2009bk}, HS1, HS2, {Haba:2011fn}}. 
Since the 
 neutrino Yukawa couplings are not necessarily tiny anymore,
 some related researches have been done, such as,  
 collider phenomenology\cite{Davidson:2010sf,HabaTsumura}, 
 low energy thermal leptogenesis\cite{HS1, HS2},  
 cosmological constraints\cite{Sher:2011mx}\footnote{
A setup in Refs.\cite{Sher:2011mx} is different from 
 usual neutrinophilic Higgs doublet models, since 
 it includes a light Higgs particle.  
 }, and so on.

On the other hand, 
 for the SUSY, 
 it is the most promising candidate 
 as a new physics beyond the SM  
 because of a excellent success of gauge coupling unification. 
Thus, the SUSY SM well fits the GUT scenario as 
 well as an existence of 
 a dark matter candidate.

There are some attempts that try to realize 
 the relation in 
 Eq.(\ref{1}).
One example is 
 to derive $m_\nu$ from a higher dimensional 
 operator in the SUSY
 framework\cite{ArkaniHamed:2000bq}. 
Another example is to take a setup of 
 matter localization\cite{Agashe:2008fe}  
 in a warped extra dimension\cite{RS}. 
Both scenarios are interesting, but 
 a model in this letter is much simpler 
 and contains no additional scales other than 
 $M_{NP}$, $m_\nu$, and $M_{GUT}$.     
(For other related papers, see, for example, 
 \cite{{related},{related2}}.)


In this letter, 
 we suggest 
 a simple SUSY neutrinophilic Higgs doublet model, which 
 dynamically realizes the relation 
 of Eq.(\ref{1}).  
Usually,
 SUSY neutrinophilic doublet models have 
 tiny mass scale of 
 soft $Z_2$-symmetry breaking  
 ($\rho,\rho'={\cal O}(10)$ eV   
 in Refs.\cite{HS1,HS2,{Haba:2011fn}}). 
This additional tiny mass scale 
 plays a crucial role of 
 generating the tiny neutrino mass, 
 however, 
 its origin 
 is completely unknown (and assumption). 
In other words, 
 the smallness of $m_\nu$ is just replaced by 
 that of $Z_2$-symmetry breaking mass parameters, 
 and this is not an essential explanation of 
 tiny $m_\nu$.   
This is a common serious problem exists in 
 neutrinophilic Higgs doublet models in general. 
%
%
Our model solves this problem, where 
 two scales of
 $M_{GUT}$ and $M_{NP}$  
 naturally induce  
 the suitable magnitude of $m_\nu$  
 through the relation of Eq.(\ref{1}),   
 and does not require 
 any additional scales.  
The model contains a pair of new neutrinophilic Higgs doublets
 with GUT-scale masses, and 
 the $Z_2$-symmetry is broken by TeV-scale 
 dimensionful couplings
 of these new doublets to the ordinary SUSY
 Higgs doublets. 
Once the ordinary Higgs doublets obtain VEVs $(v_{u,d})$ by the usual
 electroweak symmetry breaking,
 it triggers VEVs for the neutrinophilic 
 Higgs doublets of size,
 $m_\nu \sim v_{u,d}M_{NP}/M_{GUT}$. 
Then, ${\cal O}(1)$ Yukawa
 couplings of the neutrinophilic doublets to
 $L\ N$
 ($L$: lepton doublet, $N$: right-handed neutrino)  
 give neutrino
 masses of the proper size.
A gauge coupling unification 
 is also preserved automatically 
 in our 
 setup.

\section{SUSY neutrinophilic Higgs doublet model}

At first, we show a SUSY neutrinophilic Higgs doublet 
 model 
 in a parameter region which is different from
 Refs.\cite{HS1, HS2, {Haba:2011fn}}. 
We introduce $Z_2$-parity, where 
 only vector-like neutrinophilic Higgs doublets and 
 right-handed neutrino have odd-charge. 
The superpotential of the Higgs sector 
 is given by\footnote{
The author would like to thank R. Kitano 
 for pointing out a paper\cite{Kitano:2002px}, which 
 suggested the similar model and also estimated 
 lepton flavor violating processes.   
} 
\begin{eqnarray}
\mathcal{W}_h = 
 \mu H_u H_d  + M H_{\nu} H_{\nu'} - \rho H_u H_{\nu'} - \rho' H_{\nu} H_d.
\label{WW}
\end{eqnarray}
$H_\nu$ ($H_{\nu'}$) is a neutrinophilic 
 Higgs doublet, 
 and $H_\nu$ has Yukawa interaction 
 of $LH_\nu N$, which induces a tiny Dirac neutrino 
 mass through the tiny VEV,  
 $\langle H_\nu\rangle$.  
This is the origin of smallness of 
 neutrino mass,  
 and this paper devotes a Dirac neutrino scenario, i.e., 
 $m_\nu \simeq \langle H_\nu\rangle ={\cal O}(0.1)$ eV.  
On the other hand, 
 $H_{\nu'}$ does not couple with any matters. 
$H_u$ and $H_d$ are 
 Higgs doublets in 
 the minimal SUSY SM 
 (MSSM), and  
 quarks and charged lepton obtain their masses 
 through 
 $\langle H_u\rangle$ and $\langle H_d\rangle$. 
Note that this structure is guaranteed by 
 the $Z_2$-symmetry.  
Differently from conventional 
 neutrinophilic Higgs doublet models, 
 we here take $M$ the GUT scale  
 and $\mu, \rho, \rho'$  
 ${\cal O} (1)$ TeV. 
The soft $Z_2$-parity breaking parameters, 
 $\rho$ and $\rho'$, might be  
 induced from SUSY breaking effects  
 (which will be discussed in the next section),
 and 
 we 
 regard   
 $\rho$ and $\rho'$ as 
 mass 
 parameters of  
 new physics scale, $M_{NP}={\cal O}(1)$ TeV. 
Usually, 
 SUSY neutrinophilic doublet models take 
 $\rho,\rho'={\cal O}(10)$ eV 
 (for ${\cal O}(1)$ TeV $B$-terms)\cite{HS1, HS2, {Haba:2011fn}}. 
This additional tiny mass scale plays a crucial role of 
 generating the tiny neutrino mass  
 however, 
 its origin 
 is just an assumption. 
Thus, 
 the smallness of $m_\nu$ is just replaced by 
 that of $\rho, \rho'$.  
This is a common serious problem exists in 
 neutrinophilic Higgs doublet models in general. 
The present model solves this problem, in which 
 two scales of
 $M_{GUT}$ and $M_{NP}$  
 induce  
 the suitable magnitude of $m_\nu$
 dynamically,  
 and does not require 
 any additional scales, such as 
 ${\cal O}(10)$ eV. 
This is one of the excellent points 
 in our model.

The potential of the Higgs doublets 
 is given by 
\begin{eqnarray}
 V &=& (|\mu|^2 +|\rho|^2) H_u^\dag H_u + (|\mu|^2+|\rho'|^2) H_d^\dag H_d 
      + (|M|^2 +|\rho'|^2) H_{\nu}^\dag H_{\nu}
  + (|M|^2+|\rho|^2) H_{\nu'}^\dag H_{\nu'}  \nonumber \\
  && + \frac{g_1^2}{2} \left( H_u^\dag \frac{1}{2} H_u
     - H_d^\dag\frac{1}{2} H_d 
     + H_{\nu}^\dag \frac{1}{2} H_{\nu}
     - H_{\nu'}^\dag \frac{1}{2}H_{\nu'} \right)^2  \nonumber \\
  && + \sum_a \frac{g_2^2}{2} \left( H_u^\dag \frac{\tau^a}{2} H_u
     + H_d^\dag\frac{\tau^a}{2} H_d 
     + H_{\nu}^\dag \frac{\tau^a}{2} H_{\nu}
    + H_{\nu'}^\dag \frac{\tau^a}{2}H_{\nu'} \right)^2  \nonumber \\
  && - m_{H_u}^2 H_u^\dag H_u  + m_{H_d}^2 H_d^\dag H_d 
     + m_{H_\nu}^2 H_{\nu}^\dag H_{\nu} 
     + m_{H_{\nu'}}^2 H_{\nu'}^\dag H_{\nu'} \nonumber \\
  && + B \mu H_u \cdot H_d + B' M H_{\nu}\cdot H_{\nu'}
     - \hat{B} \rho H_u \cdot H_{\nu'}
     - \hat{B}' \rho' H_{\nu}\cdot H_{d}\nonumber\\
&& - \mu^* \rho H_d^\dag H_{\nu'}
   -\mu^* \rho' H_u^\dag H_{\nu}
   - M^* \rho' H_{\nu'}^\dag H_{d}
   - M^* \rho H_\nu^\dag H_{u}
 + \text{h.c.} ,
\end{eqnarray}
where $\tau^a$ and dot mean 
 a generator and cross product of $SU(2)$, respectively. 
$m_{H_u}^2$, $m_{H_d}^2$, $m_{H_\nu}^2$,
 $m_{H_\nu'}^2$, $B$, $B'$, $\hat{B}$, and 
 $\hat{B}'$ are soft SUSY breaking 
 parameters of order ${\cal O}(1)$ TeV. 
We assume $(-m^2_{H_u})<0$ for the suitable electroweak symmetry
 breaking,  
 and real VEVs as    
 $\langle H_u \rangle = v_u$, 
 $\langle H_d \rangle = v_d$, 
 $\langle H_\nu \rangle = v_\nu$,  
 $\langle H_{\nu'} \rangle = v_{\nu'}$ 
 in neutral components.  
Now let us examine whether we can really obtain 
 the suitable magnitudes of VEVs  
 as $v_{u,d}={\cal O}(10^2)$ GeV and 
 $v_{\nu, \nu'}={\cal O}(0.1)$ eV or not. 
By taking $\mu$-, $\rho$-, $B$-parameters to be real,   
 and denoting 
 $M_u^2 \equiv \mu^2 +\rho^2- m_{H_u}^2 (<0)$, 
 $M_d^2 \equiv \mu^2 +\rho'^2+ m_{H_d}^2 (>0)$, 
 $M_\nu^2 \equiv M^2 +\rho'^2- m_{H_u}^2 \simeq M^2 (>0)$, and 
 $M_{\nu'}^2 \equiv M^2 +\rho^2+ m_{H_d}^2 \simeq M^2 (>0)$, 
 the stationary conditions of 
 $\frac{1}{2} \frac{\partial V}{\partial v_u}=0$, 
 $\frac{1}{2} \frac{\partial V}{\partial v_d}=0$,
 $\frac{1}{2} \frac{\partial V}{\partial v_\nu}=0$,
 $\frac{1}{2} \frac{\partial V}{\partial v_{\nu'}}=0$    
 are given by 
\begin{eqnarray}
&&\hspace*{-5mm}
0 
   = M^2_u v_u + \frac{1}{4} (g_1^2 + g_2^2) v_u (v_u^2 - v_d^2 
     + v_{\nu}^2 - v_{\nu'}^2) + B \mu v_d -  \hat{B} \rho v_{\nu'} - (
     \mu \rho' + M \rho ) v_{\nu}, 
\label{ssc1} \\ 
&&\hspace*{-5mm}
0 
   = M^2_d v_d - \frac{1}{4}(g_1^2 + g_2^2) v_d (v_u^2 - v_d^2 
     + v_{\nu}^2 - v_{\nu'}^2) +  B \mu v_u -  \hat{B}' \rho' v_{\nu} 
     - ( \mu \rho +  M \rho' ) v_{\nu'} ,
\label{ssc2} \\ 
&&\hspace*{-5mm}
0 
   = M_{\nu}^2 v_{\nu} + \frac{1}{4}(g_1^2 + g_2^2) v_{\nu}
    (v_u^2 - v_d^2  + v_{\nu}^2 - v_{\nu'}^2) 
   +  B' M v_{\nu'} -  \hat{B}' \rho' v_d 
   - (\mu \rho' +  M \rho ) v_u ,
\label{ssc3} \\ 
&&\hspace*{-5mm}
0 
   = M_{\nu'}^2 v_{\nu'} - \frac{1}{4}(g_1^2 + g_2^2) v_{\nu'} (v_u^2 - v_d^2 
     + v_{\nu}^2 - v_{\nu'}^2) +  B' M v_{\nu}
   - \hat{B} \rho v_u - ( \mu \rho + M \rho' ) v_d , 
\label{ssc4} 
\end{eqnarray}
respectively. 
Regarding $M$ 
 is a GUT scale and 
 $v_\nu, v_{\nu'} \ll v_u, v_d$, 
 Eqs.(\ref{ssc3}) and (\ref{ssc4}) 
 become 
\begin{eqnarray}
0  = M v_{\nu} -  \rho  v_u , \;\;\;
0  = M v_{\nu'} - \rho' v_d 
\label{rel2}
\end{eqnarray}
in the leading order, respectively. 
These are just 
 the relation in Eq.(\ref{1}) ! 
(Here we neglect one order magnitude 
 between $M_{NP}$ and the weak scale.)  
This is what we want to derive, and 
 the VEVs of 
 neutrinophilic Higgs fields become  
\begin{eqnarray}
v_{\nu} =  \frac{\rho  v_u}{M} , \;\;\;
v_{\nu'} = \frac{\rho' v_d}{M} .  
\label{wvev}
\end{eqnarray}
It is worth noting that 
 they are induced dynamically through the 
 stationary conditions in Eqs.(\ref{ssc3}) and (\ref{ssc4}),   
 and their magnitudes are surely 
 of ${\cal O}(0.1)$ eV.  
As for 
 $v_{u,d}$, 
 by use of Eq.(\ref{wvev}),   
 Eqs.(\ref{ssc1}) and (\ref{ssc2}) 
 become  
\begin{eqnarray}
&&
0  = (M^2_u-\rho^2) v_u + \frac{1}{4} (g_1^2 + g_2^2) v_u (v_u^2 - v_d^2 
     + v_{\nu}^2 - v_{\nu'}^2) + B \mu v_d , \\
&& 
0  = (M^2_d-\rho'^2) v_d - \frac{1}{4}(g_1^2 + g_2^2) v_d (v_u^2 - v_d^2 
     + v_{\nu}^2 - v_{\nu'}^2) +  B \mu v_u 
\end{eqnarray}
in the leading order, respectively.  
Then, the MSSM Higgs fields take VEVs as 
\begin{eqnarray}
&& v^2 \simeq  \frac{2}{g_1^2 + g_2^2} \left( \frac{M_u'^2 -
					M_d'^2}{\cos{2\beta}} -(M_u'^2 +
					M_d'^2) \right) , \;\;\;\;\;
 \sin{2\beta} \simeq  \frac{2B \mu}{M_u'^2 + M_d'^2 },  
\end{eqnarray}
where 
 $v^2=v_u^2+v_d^2$, $\tan\beta =v_u/v_d$, 
 $M_u'^2\equiv M^2_u-\rho^2$ and 
 $M_d'^2\equiv M^2_d-\rho'^2$. 
They mean  
 slight modifications of VEVs for 
 $H_u$ and $H_d$.

Since the masses of neutrinophilic Higgs $H_\nu$ and 
 ${H}_{\nu'}$ are super-heavy as the GUT scale, 
 there are no other vacua (such as,  
 $v_{u,d} \sim v_{\nu, \nu'}$)
 except for 
 $v_{u,d} \gg v_{\nu, \nu'}$ \cite{Haba:2011fn}. 
Also, 
 their heaviness guarantees the stability 
 of the VEV hierarchy, 
 $v_{u,d} \gg v_{\nu, \nu'}$, against radiative 
 corrections \cite{Morozumi, Haba:2011fn}. 
It is because, 
 in the effective potential,  
 $H_\nu$ and 
 ${H}_{\nu'}$ inside 
 loop-diagrams are suppressed 
 by their GUT scale masses. 
Anyhow, 
 we stress again that 
 the relation of Eq.(\ref{1}) is 
 dynamically obtained in 
 Eq.(\ref{rel2}). 

As for the
 gauge coupling unification, 
 it is preserved automatically 
 in our
 setup, since 
 fields except for 
 the MSSM have masses of 
 order the GUT scale.

\section{$SU(5)$ GUT embedded model}

The model we suggested 
 has 
 the GUT scale mass of neutrinophilic Higgs doublets 
 in Eq.(\ref{WW}), 
 so that it is natural to embed the model 
 in a GUT framework. 
Let us consider $SU(5)$ GUT. 
A superpotential of a Higgs sector 
 at the GUT scale  
 is given by
\begin{eqnarray}
\mathcal{W}_H^{\rm GUT} = 
 M_0 {\rm tr}\Sigma^2 + \lambda {\rm tr}\Sigma^3 + 
 H \Sigma \bar{H} + \Phi_\nu \Sigma \bar{\Phi}_\nu
 - M_1 H \bar{H}  - M_2 \Phi_\nu \bar{\Phi}_\nu  ,
\label{WGUT}
\end{eqnarray}
where $\Sigma$ is an adjoint Higgs whose VEV reduces the 
 GUT gauge symmetry into the SM. 
$\Phi_\nu$ ($\bar{\Phi}_\nu$) is a neutrinophilic 
 Higgs of (anti-)fundamental 
 representation, which contains $H_\nu$ ($H'_\nu$) 
 in the doublet component    
 (while the triplet component is denoted as $T_\nu$ ($\bar{T}_\nu$)). 
$\Phi_\nu$ and $\bar{\Phi}_\nu$ are odd under 
 the 
 $Z_2$-parity. 
$H$ ($\bar{H}$) is a 
 Higgs of (anti-)fundamental 
 representation, which contains $H_u$ ($H_d$) 
 in the doublet component 
 (while the triplet component is denoted as $T$ ($\bar{T}$)). 
The VEV of $\Sigma$ and $M_{0,1,2}$ are all  
 of ${\cal O}(10^{16})$ GeV, 
 thus we encounter so-called triplet-doublet (TD)
 splitting problem. 
Some mechanisms have been suggested 
 for a solution of TD splitting, 
 but here 
 we show a case that 
 the TD splitting is realized 
 just by a fine-tuning 
 between $\langle \Sigma \rangle$ and $M_1$. 
That is, 
 $\langle \Sigma \rangle - M_1$ induces 
 GUT scale masses of $T, \bar{T}$, 
 while 
 weak scale masses of $H_u, H_d$. 
This is a serious fine-tuning, 
 so that we can not expect a simultaneous 
 fine-tuned cancellation also happens 
 between 
 $\langle \Sigma \rangle$ and $M_2$.  
Thus, we 
 consider a case that the TD splitting only 
 works in 
 $H$ and $\bar{H}$, 
 while not works in 
 $\Phi_\nu$ and $\bar{\Phi}_\nu$. 
This situation makes Eq.(\ref{WGUT}) become 
%
\begin{eqnarray}
\mathcal{W}^{eff}_H = 
 \mu H_u H_d 
 + M H_{\nu} H_{\nu'} 
 + M'T \bar{T}
 + M'' T_\nu \bar{T}_\nu.
\label{WGUT2}
\end{eqnarray}
This is the effective superpotential of the Higgs sector  
 below the GUT scale, and   
 $M, M', M''$ are of ${\cal O}(10^{16})$ GeV, while 
 $\mu={\cal O} (1)$ TeV.


Now let us consider 
 an origin of  
 soft $Z_2$-parity breaking 
 terms, $\rho H_u H_{\nu'}$ and $\rho' H_{\nu} H_d$ in 
 Eq.(\ref{WW}). 
They play a crucial role of 
 generating
 the marvelous relation in Eq.(\ref{1})
 as well as 
 a tiny Dirac neutrino mass. 
Since the values of 
 $\rho, \rho'$ are 
 of order ${\cal O}(1)$ TeV, 
 they might be induced from 
 the SUSY breaking effects. 
We can consider some 
 possibilities for this mechanism. 
One example is to 
 introduce  
 a singlet $S$ with odd $Z_2$-parity. 
The superpotential including $S$ 
 below the GUT scale 
 is given by 
\begin{eqnarray}
\mathcal{W}^{eff}_S &=& 
 \mu H_u H_d 
 + M H_{\nu} H_{\nu'} 
 + M'T \bar{T}
 + M'' T_\nu \bar{T}_\nu \nonumber \\ 
&& + \mu_S S^2 + \frac{1}{\Lambda} S^4
  - SH_uH_{\nu'} - S H_\nu H_d 
  - ST\bar{T}_{\nu} - S T_\nu \bar{T} .
\label{WGUT3}
\end{eqnarray} 
Denoting $\langle T \rangle =t$, 
 $\langle \bar{T} \rangle =\bar{t}$, 
 $\langle T_\nu \rangle =t_\nu$, 
 $\langle \bar{T}_\nu \rangle =\bar{t}_\nu$, and 
 $\langle S \rangle =s$, 
 the effective potential of the Higgs sector 
 is given by 
\begin{eqnarray}
 V_S^{eff} &=& 
 |M v_\nu - sv_u|^2 + 
 |M v_{\nu'} - sv_d|^2 + 
 |M'' t_\nu - st|^2 + 
 |M'' \bar{t}_{\nu} - s\bar{t}|^2 \nonumber \\
  &&\hspace*{-3mm} + 
 |\mu v_d - sv_{\nu'}|^2 + 
 |\mu v_u - sv_\nu|^2 + 
 |M' t - st_\nu|^2 + 
 |M' \bar{t} - s\bar{t}_\nu|^2 \nonumber \\
  &&
\hspace*{-3mm} + 
 |2\mu_S s + 4s^3/\Lambda -v_u v_{\nu'} -v_\nu v_d
  -t\bar{t}- t_\nu\bar{t}_\nu|^2 
\label{Spot} \\
  &&
\hspace*{-5mm} - m_{S}^2 s^2
     - m_{H_u}^2 v_u^2
     + m_{H_d}^2 v_d^2 
     + m_{H_\nu}^2 v_{\nu}^2 
     + m_{H_{\nu'}}^2 v_{\nu'}^2 
     + m_{T}^2 t^2 
     + m_{\bar{T}}^2 \bar{t}^2 
     + m_{T_\nu}^2 t_{\nu}^2 
     + m_{\bar{T}_{\nu}}^2 \bar{t}_{\nu}^2 +\cdots \ , \nonumber 
\end{eqnarray}
where we omit $D$- and $B$-terms 
 for simplicity. 
The last line in Eq.(\ref{Spot}) are
 soft SUSY breaking mass squared terms. 
Taking a parameter region of 
 $-m_S'^2\equiv -m_S^2+4\mu_S^2 <0$, 
 we obtain 
 $s \simeq \sqrt{{\Lambda}/({32\mu_S})}\ m_S'$. 
Then, 
 when 
 $m_S'\sim \mu_S\sim 1$ TeV and  
 $\Lambda \sim 30$ TeV, 
 the suitable 
 $\rho$- and $\rho'$-terms in Eq.(\ref{WW}) 
 are effectively induced
 through 
 $s\sim 1$ TeV.  
This vacuum also suggests 
 the suitable VEVs of 
 $v_{u,d}\sim 100$ GeV, $v_{\nu,\nu'}\sim 0.1$ eV as well as  
 $t=\bar{t}=t_\nu=\bar{t}_\nu=0$.  
Unfortunately, 
 the scale of $\Lambda \sim 30$ TeV 
 is a little artificial 
 in this example. 
But, this is around $M_{NP}$, 
 and much better than inputting 
 ${\cal O}(10)$ eV as the $Z_2$-parity breaking soft mass 
 parameters. 
%
Another example is to 
 take a non-canonical K$\ddot{\rm a}$hler of  
 $[S^\dagger (H_uH_{\nu'} + H_\nu H_d)+{\rm h.c.}]_D$. 
Where $F$-term of $S$ could induce 
 the $\rho$- and $\rho'$-terms effectively 
 through the SUSY breaking scale 
 as in 
 Giudice-Masiero mechanism\cite{GM}.   
There might be other models 
 which induce the $\rho$- and $\rho'$-terms
 in Eq.(\ref{WW}) except for 
 introducing a singlet $S$.  

\section{Summary}

Among three typical energy scales, 
 a neutrino mass scale, 
 a GUT scale, 
 and  
 a TeV (SUSY)-scale, 
 there is a marvelous relation of Eq.(\ref{1}). 
In this paper, 
 we have suggested  
 a simple supersymmetric neutrinophilic 
 Higgs doublet model, which 
 realizes the relation 
 of Eq.(\ref{1})  
 dynamically as well as  
 the suitable $m_\nu$ through  
 a tiny VEV 
 of neutrinophilic Higgs  
 without 
 additional scales other than 
 $M_{NP}$ and $M_{GUT}$.     
Usually,
 SUSY neutrinophilic doublet models have 
 tiny mass scale of 
 soft $Z_2$-symmetry breaking as 
 $\rho,\rho'={\cal O}(10)$ eV.  
This additional tiny mass scale plays a crucial role of 
 generating the tiny neutrino mass, 
 however, 
 its origin 
 is just an assumption. 
In other words, 
 the smallness of $m_\nu$ is just replaced by 
 that of $Z_2$-symmetry breaking mass parameters, 
 and this is not an essential explanation of 
 tiny $m_\nu$.   
This is a common serious problem exists in 
 neutrinophilic Higgs doublet models in general. 
Our model have solved this problem, where 
 two scales of
 $M_{GUT}$ and $M_{NP}$  
 naturally induce  
 the suitable magnitude of $m_\nu$  
 through the relation of Eq.(\ref{1}),   
 and does not require 
 any additional scales.  
A gauge coupling unification 
 is also preserved automatically 
 in our
 setup. 
We have also considered the embedding in $SU(5)$ GUT 
 and some candidates of inducing 
 the $Z_2$-symmetry breaking terms from the 
 SUSY breaking effects.


\vspace{1cm}

{\large \bf Acknowledgments}\\

\vspace{-.2cm}
\noindent
We thank T. Horita 
 for useful discussions. 
This work is partially supported by Scientific Grant by Ministry of 
 Education and Science, Nos. 20540272, 20039006, and 20025004.


\end{document}